\documentclass[preprint]{aastex62}
\usepackage{verbatim}

\received{}
\revised{}
\accepted{}

\submitjournal{ApJ}

\shorttitle{A Proposed Dual AGN with the EVN}
\shortauthors{P.~M. Veres et al.}

\begin{document}

\title{European VLBI Network Observations of the Proposed Dual AGN SDSS\,J101022.95$+$141300.9}

\author[0000-0002-9553-2987]{Patrik Mil\'an Veres}
\affiliation{ELTE E\"otv\"os Lor\'and University, Institute of Geography and Earth Sciences, Department of Astronomy, P\'azm\'any P\'eter s\'et\'any 1/A,
H-1117 Budapest, Hungary}
\affiliation{Konkoly Observatory, ELKH Research Centre for Astronomy and Earth Sciences, Konkoly Thege Mikl\'os \'ut 15-17, H-1121 Budapest, Hungary}

\author[0000-0003-1020-1597]{Krisztina \'Eva Gab\'anyi}
\affiliation{ELTE E\"otv\"os Lor\'and University, Institute of Geography and Earth Sciences, Department of Astronomy, P\'azm\'any P\'eter s\'et\'any 1/A,
H-1117 Budapest, Hungary}
\affiliation{ELKH-ELTE Extragalactic Astrophysics Research Group, ELTE E\"otv\"os Lor\'and University, P\'azm\'any P\'eter s\'et\'any 1/A, H-1117 Budapest, Hungary}
\affiliation{Konkoly Observatory, ELKH Research Centre for Astronomy and Earth Sciences, Konkoly Thege Mikl\'os \'ut 15-17, H-1121 Budapest, Hungary}

\author[0000-0003-3079-1889]{S\'andor Frey}
\affiliation{Konkoly Observatory, ELKH Research Centre for Astronomy and Earth Sciences, Konkoly Thege Mikl\'os \'ut 15-17, H-1121 Budapest, Hungary}
\affiliation{Institute of Physics, ELTE E\"otv\"os Lor\'and University,
P\'azm\'any P\'eter s\'et\'any 1/A,
H-1117 Budapest, Hungary}

\author[0000-0002-5195-335X]{Zsolt Paragi}
\affiliation{Joint Institute for VLBI ERIC, Oude Hoogeveensedijk 4, 7991 PD Dwingeloo, The Netherlands}

\author[0000-0003-2769-3591]{Emma Kun}
\affiliation{Konkoly Observatory, ELKH Research Centre for Astronomy and Earth Sciences, Konkoly Thege Mikl\'os \'ut 15-17, H-1121 Budapest, Hungary}

\author[0000-0003-4341-0029]{Tao An}
\affiliation{Shanghai Astronomical Observatory, Key Laboratory of Radio Astronomy, Chinese Academy of Sciences, 80 Nandan Road, Shanghai 200030, China}

\correspondingauthor{Krisztina \'Eva Gab\'anyi}
\email{krisztina.g@gmail.com}

\begin{abstract}

During galaxy merger events, the supermassive black holes in the center of the galaxies may form a pair of active galactic nuclei (AGN) with kpc-scale or even pc-scale separation. Recently, optical observations revealed a promising dual AGN candidate at the center of the galaxy SDSS\,J101022.95$+$141300.9 (hereafter J1010$+$1413). The presence of two distinct [\ion{O}{3}]-emitting point sources with a projected separation of $\sim 430$\,pc indicates a dual AGN system. To search for AGN-dominated radio emission originating from the \textit{Hubble Space Telescope} (\textit{HST}) point sources, we carried out very long baseline interferometry observations. We resolved the radio structure of J1010$+$1413 and detected a single feature offset from the \textit{HST} point sources and also from the \textit{Gaia} optical position of the object. Our multi-wavelength analysis of J1010$+$1413 inferred two possible interpretations of the observed properties challenging its proposed dual AGN classification.

\end{abstract}


\section{Introduction} 
\label{sec:intro}

Dual active galactic nuclei (AGN) are powered by supermassive black holes (SMBHs) hosted in merging galaxies \citep{binaries}. These systems are characterized by accreting SMBHs in a wide range of projected separations, between pc and $\sim 100$\,kpc scales \citep{multi-messenger}. At smaller (sub-pc) separations, gravitationally bound binary SMBHs are on the way to eventually merge, becoming potential sources of nanohertz gravitational waves \citep[GWs;][]{GW} that are expected to be detected with Pulsar Timing Arrays (PTAs) in the near future \citep{PTA}. Observational studies of dual and binary AGN are also important for understanding the AGN triggering mechanism during galaxy mergers, and providing inputs for simulations of the galaxy and SMBH evolution throughout the history of the Universe \citep{multi-messenger}.

Optical spectroscopy offers a means of detecting dual AGN since double-peaked emission lines may indicate the presence of two accreting AGN with different radial velocity components in a merging system (e.g., \citealp{double-peaked}). However, these double-peaked lines may often be signs of activity in a single AGN, like AGN-driven outflows \citep[e.g.,][]{multi-messenger} and/or rotating narrow-line region \citep[NLR; e.g.,][and references therein]{nlr}.
Consequently, further observations at other wavelengths are required to strengthen or falsify the dual AGN identification. Radio observations at cm wavelengths are able to confirm the existence of jetted AGN at the center of massive galaxies, since the synchrotron radiation emitted by the jets is detectable with milli-arcsec (mas) resolution using the very long baseline interferometry (VLBI) technique. Moreover, VLBI is capable of spatially resolving dual radio-emitting AGN, thus providing direct imaging evidence for duality \citep{dual-radio}. However, only about $10\%$ of AGN are radio-loud \citep{radio-loud}, hence VLBI observations alone cannot completely reject the possibility of the presence of two AGN in a system. Nevertheless, radio interferometric observations were essential elements of identifying dual AGN in several cases, e.g., 3C\,75 \citep[][and references therein]{3C75, 3C75(2)} and SDSS\,J1502+1115
\citep{sdssdualagn,deane-triple}.

Recently, \citet{Goulding} identified two compact [\ion{O}{3}]-emitting regions at the center of the galaxy SDSS\,J101022.95$+$141300.9 (hereafter J1010$+$1413). From the results of the Wide Field Camera 3 (WFC3) observations onboard the \textit{Hubble Space Telescope} (\textit{HST}), they concluded that the optical signatures of J1010$+$1413 suggest the presence of a dual AGN system\footnote{\citet{Goulding} call this system binary but we follow the conventional terminology \citep{multi-messenger} and, based on the proposed SMBH separation, designate it as a dual AGN candidate.} with a projected separation of $\sim 430$\,pc. The F689M filter image shows two clearly resolved distinct stellar-continuum point sources which are spatially coincident with two [\ion{O}{3}]-emitting point sources. According to \citet{Goulding}, these indicate the presence of two NLRs (i.e., a dual AGN system). 

Earlier \citet{Jarvis} observed ten low-redshift quasars including J1010$+$1413 with the Karl G. Jansky Very Large Array (VLA). Their radio images taken at three different frequencies show a structure extended beyond the [\ion{O}{3}]-emitting region. However, their data cannot be applied to check the duality, since the VLA angular resolution was not sufficient to distinguish between the possible two [\ion{O}{3}]-emitting point sources found by \citet{Goulding}. To search for radio emission inside the [\ion{O}{3}]-emitting region and to resolve the radio structure detected by the VLA, we carried out high-resolution radio interferometric observations with the European VLBI Network (EVN) and the enhanced Multi-Element Remotely Linked Interferometer Network (e-MERLIN). 

In this work, we present the results of our VLBI observations of the dual AGN candidate J1010$+$1413 taken at two frequencies. The observations and data analysis are described in Section~\ref{sec:obs}. The results are presented in Section~\ref{sec:res} and discussed in Section~\ref{sec:disc}. We interpret the available multi-band observing data in the context of a single AGN and a dual AGN scenario in Section~\ref{sec:scenarios}. Conclusions are given in Section~\ref{sec:concl}.
Throughout this paper, we adopt a flat $\Lambda$CDM cosmological model with H$_{0}=70$\,km s$^{-1}$ Mpc$^{-1}$, $\Omega_{\Lambda}=0.73$, and $\Omega_{\mathrm{m}}=0.27$. In this model \citep{cosmology}, $1$\,mas angular size corresponds to $3.3$\,pc projected linear size at the source redshift $z=0.199$ \citep{redshift}.

\section{Observations and Data Analysis} 
\label{sec:obs}

We observed J1010$+$1413 with the EVN and e-MERLIN. The observations were carried out at $1.7$\,GHz on 2020 March 12 and at $4.9$\,GHz on 2020 February 26 (project code: EG109, PI: K. Gab\'anyi). Thirty-two and sixty-four 500-kHz spectral channels were included in each of the eight intermediate frequency channels (IFs) resulting in total bandwidths of 128 MHz and 256 MHz for the $1.7$\,GHz and $4.9$\,GHz observations, respectively. Further information is given in Table \ref{tab:obs}. The experiments were performed in phase-referencing mode and lasted for $3$\,h (at 1.7\,GHz) and $2$\,h (at 4.9\,GHz). The data were recorded at $1$~Gbit~s$^{-1}$ (at 1.7\,GHz) and $2$~Gbit~s$^{-1}$ maximum rate (at 4.9\,GHz)
in left and right circular polarizations and correlated with 2~s averaging time at the EVN Data Processor \citep{JIVEcorrelator} at the Joint Institute for VLBI European Research Infrastructure Consortium (Dwingeloo, The Netherlands). Two calibrator sources were observed, SDSS\,J100741.49$+$135629.6 (hereafter J1007+1356) as phase reference and SDSS\,J095649.88$+$251516.0 for fringe finding. The phase-reference calibrator J1007$+$1356 located $0\fdg71$ apart from J1010$+$1413 has an accurate radio position in the current 3rd realization of the International Celestial Reference Frame (ICRF3, \citealt{icrf}) at right ascension $10^\mathrm{h} 07^\mathrm{min} 41\fs498089 \pm 0\fs000002$ and declination $+13^{\circ} 56\arcmin 29\farcs60077 \pm 0\farcs00004$. This calibrator source has a prominent 10-mas scale inner jet pointing towards the southeast \citep{pref-jet}.

\begin{deluxetable*}{cccccc}[h!]
\tablecaption{Information about the VLBI observations of J1010$+$1413.\label{tab:obs}}
\tablecolumns{5}
\tablenum{1}
\tablewidth{0pt}
\tablehead{
\colhead{Epoch} &
\colhead{On-source time} &
\colhead{Frequency} & \colhead{Participating antennas} \\
 \colhead{(yr)} & \colhead{(min)} &
\colhead{(GHz)} & \colhead{} 
}
\startdata
$2020.194$ & $159$ & $1.7$& Wb, Ef, T6, Tr, Hh, Sv, Zc, Bd, Sr, Cm, Da, De, Pi \\
$2020.153$ & $102$ & $4.9$ & Jb, Wb, Ef, O8, T6, Tr, Ys, Hh, Sv, Zc, Bd, Ir, Km, Da \\
\enddata
\tablecomments{Bd: Badary, Russia; Cm: Cambridge, UK; Da: Darnhall, UK; De: Defford, UK; Ef: Effelsberg, Germany; Hh: Hartebeesthoek, South Africa; Ir: Irebene, Latvia; Jb: Jodrell Bank, UK; Km: Kunming, China; O8: Onsala, Sweden; Pi: Pickmere, UK; Sr: Sardinia, Italy; Sv: Svetloe, Russia; T6: Tianma, China; Tr: Toru\'n, Poland; Wb: Westerbork, The Netherlands; Ys: Yebes, Spain; Zc: Zelenchukskaya, Russia}
\end{deluxetable*}

The data were calibrated with the U.S. National Radio Astronomy Observatory (NRAO) Astronomical Image Processing System \citep[\textsc{AIPS},][]{aips} software package. Visibility amplitudes were calibrated using the measured system temperatures and gain curves of each telescope. We corrected the dispersive ionospheric delays based on total electron content measurements from Global Positioning System satellites. Simple bandpass corrections were made by flagging the first and last two channels per IF since these are often outliers in phase and low in amplitude. We then performed global fringe-fitting \citep{fringefit} for the phase-reference calibrator and fringe-finder sources. We imaged these with the \textsc{Difmap} software \citep{difmap} using hybrid mapping, including \textsc{clean} deconvolution \citep{hybridmapping} and phase and amplitude self-calibration \citep{selfcal}. In this process, we solved antenna-specific gain correction factors, which were then applied to all visibility data in AIPS. Fringe-fitting was repeated for the phase-reference calibrator, using the \textsc{clean} component model of its brightness distribution obtained in \textsc{Difmap}, to correct for the contribution of the source structure to the phases. Finally, we interpolated the fringe-fitting solutions to the target source (J1010$+$1413) and exported the calibrated visibility data for imaging in \textsc{Difmap}.    

Before imaging, the visibility data were averaged over 10-s time intervals. Because of the faintness of the target, the imaging was performed without any self-calibration. To represent the brightness distribution of J1010$+$1413, we fitted elliptical Gaussian model components \citep{modelfit} to the visibility data at both frequencies. To smooth the noise features in the residual image, we performed $1000$ \textsc{clean} iterations with a very small loop gain ($0.01$).

\section{Results} 
\label{sec:res}

\subsection{Radio Morphology}
\label{subsec:radio-morph}

\begin{figure}[h!]
\plotone{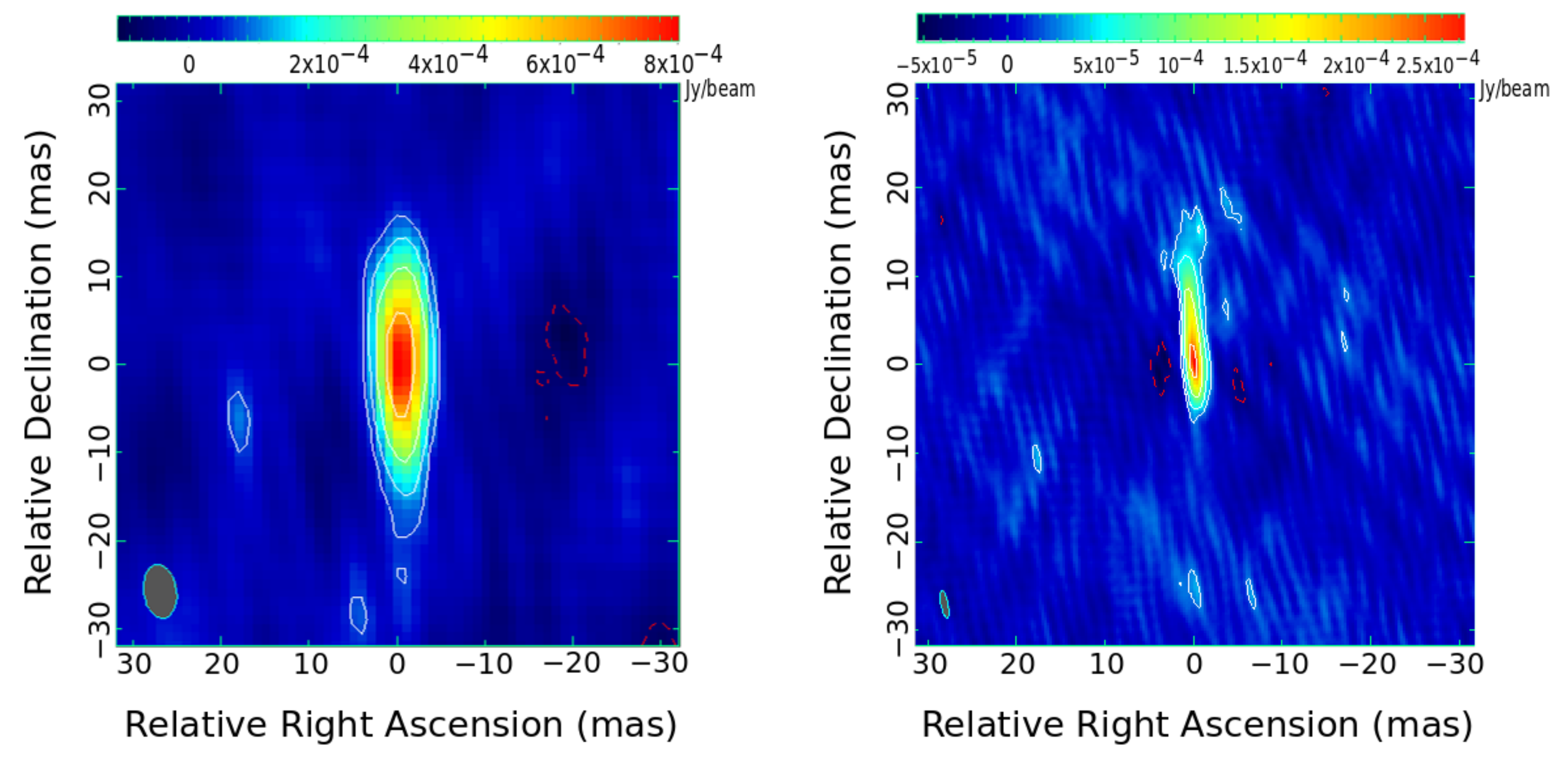}
\epsscale{1.8}
\caption{Naturally-weighted VLBI maps of J1010$+$1413 at $1.7$\,GHz (\textit{left panel}) and at $4.9$\,GHz (\textit{right panel}). In the lower-left corners, the grey ellipses represent the Gaussian restoring beams (half-power beam width, HPBW). The lowest contour levels are drawn at around $\pm 3\sigma$ image noise levels on both maps, further positive contours increase by a factor of 2. The parameters of the two maps are given in Table~\ref{tab:maps}. \label{fig:maps}}
\end{figure}

The final naturally-weighted images of J1010$+$1413 taken at the two observing frequencies are displayed in Figure~\ref{fig:maps}.
The details of the contour maps are summarized in Table~\ref{tab:maps}. Based on our EVN and e-MERLIN data, only a single feature is detected. It is elongated in the north--south direction and extended to about $40$\,mas ($\sim130$\,pc) at $1.7$\,GHz. The flux densities in the fitted Gaussian model components are $(3.33 \pm 0.33)$\,mJy at $1.7$\,GHz and $(2.05 \pm 0.21)$\,mJy at $4.9$\,GHz. Following \cite{fluxerror}, we assumed $10\%$ flux density calibration uncertainties. More details about the model components are given in Table~\ref{tab:modelfit}. 

The accuracy of the position determination depends on several factors. The absolute astrometric accuracy is affected by the positional uncertainty of the phase calibrator source and its angular separation from the target source. Furthermore, the positions of the fitted Gaussian brightness distributions are influenced by the signal-to-noise ratios of the observations, the resolution of the array, and the sizes of the fitted components. 

According to the ICRF3 \citep{icrf}, the position of the phase reference source is known with a formal error of $\sim 0.05$\,mas. However, \cite{pref-jet} showed that there is a significant frequency-dependent core shift in J1007$+$1356 in a roughly north-west to south-east direction, at a position angle of $\sim 132\degr$. The amount of core shift is $\sim0.23$\,mas between $8$\,GHz (at the frequency where the ICRF3 position is measured) and $4.9$\,GHz, and $\sim 1$\,mas between $1.7$\,GHz and $4.9$\,GHz. The $\sim 0\fdg7$ angular separation between the target and the calibrator implies a positional uncertainty of $\sim 2$\,mas at $1.7$\,GHz according to \cite{rioja}, and $\sim0.5$\,mas at $4.9$\,GHz according to \cite{Chatterjee}. Using the relations of \cite{Fomalont}, the positions of the fitted components are accurate within $\sim 0.3$\,mas and $\sim 0.2$\,mas at $1.7$\,GHz and $4.9$\,GHz, respectively. To summarize, the positional error is dominated by the value arising from the separation between the calibrator and the target source. Thus, the coordinates of J1010$+$1413 (Table~\ref{tab:modelfit}) are accurate in both right ascension and declination within the errors of $2.2$\,mas and $0.6$\,mas at $1.7$\,GHz and $4.9$\,GHz, respectively. 

\begin{deluxetable*}{ccccccccc}[h!]
\tablecaption{Details of the VLBI maps shown in Figure~\ref{fig:maps}. The last three columns give information about the elliptical Gaussian restoring beam (major and minor axes, HPBW) and the major axis position angle measured from north through east. \label{tab:maps}}
\tablecolumns{8}
\tablenum{2}
\tablewidth{0pt}
\tablehead{
\colhead{Frequency} & \colhead{Peak Brightness} & \colhead{Image Noise Level} & \colhead{Lowest Contour} & \colhead{$\theta_{\mathrm{maj}}$} & \colhead{$\theta_{\mathrm{min}}$} & \colhead{PA} \\
\colhead{(GHz)} & \colhead{(mJy beam$^{-1}$)} & \colhead{(mJy beam$^{-1}$)} & \colhead{(mJy beam$^{-1}$)} & \colhead{(mas)} & \colhead{(mas)} & ($\degr$) 
}
\startdata
$1.7$ & $0.83$ & $0.01$ & $\pm0.08$ & $6.1$ & $3.7$ & $9.0$ \\
$4.9$ & $0.28$ & $0.01$ & $\pm0.03$ & $3.1$ & $0.8$ & $10.3$ \\
\enddata
\end{deluxetable*}

\begin{deluxetable*}{cccccccc}[h!]
\tablecaption{Modelfit results. Columns (3) and (4) are the right ascension and declination of the fitted component positions. Column (5) is the FWHM size of the elliptical Gaussian major axis. Column (6) is the minor to major axis FWHM ratio. Column (7) is the major axis position angle measured from north through east. \label{tab:modelfit}}
\tablecolumns{7}
\tablenum{3}
\tablewidth{0pt}
\tablehead{
\colhead{Frequency} & \colhead{Flux Density} & \colhead{Right Ascension} & \colhead{Declination} & \colhead{FWHM} & \colhead{Axial Ratio} & \colhead{PA} \\ 
\colhead{(GHz)} & \colhead{(mJy)} & \colhead{} & \colhead{} & \colhead{(mas)} & & ($\degr$)\\
(1) & (2) & (3) & (4) & (5) & (6) & (7)
}
\startdata
$1.7$ & $3.33 \pm 0.33$ & $10^\mathrm{h} 10^\mathrm{m} 22\fs95634 \pm 0\fs00015$ & $+14\degr 13' 00\farcs8012 \pm 0\farcs0022$ & $17.46$ & $ 0.17$ & $0.1$ \\
$4.9$ & $2.05 \pm 0.21$ & $10^\mathrm{h} 10^\mathrm{m} 22\fs95650 \pm 0\fs00004$ & $+14\degr 13' 00\farcs7997 \pm 0\farcs0006$ & $12.48$ & $0.12$ & $3.3$ \\
\enddata
\end{deluxetable*}

\subsection{Brightness Temperature} 
\label{subsec:Tb}

We derived the brightness temperature ($T_{\mathrm{b}}$) of the source based on the fitted model component at 4.9\,GHz using
\begin{equation}
T_{\mathrm{b}} = 1.22 \times 10^{12} (1+z) \frac{S}{\theta_{\mathrm{maj}} \theta_{\mathrm{min}} \nu^{2}}\,\textrm{K}
\end{equation}
\citep{Tb}, where $S$ is the flux density of the Gaussian component in Jy, $\theta_{\mathrm{maj}}$ and $\theta_{\mathrm{min}}$ are the major and minor axes (full width at half-maximum, FWHM) in mas, and $\nu$ is the observing frequency in GHz. The brightness temperature obtained is $T_{\mathrm{b}} \approx 10^{7}$\,K.

\subsection{Radio Spectral Index}
\label{subsec:spx}

\begin{figure}[h!]
\epsscale{0.8}
\plotone{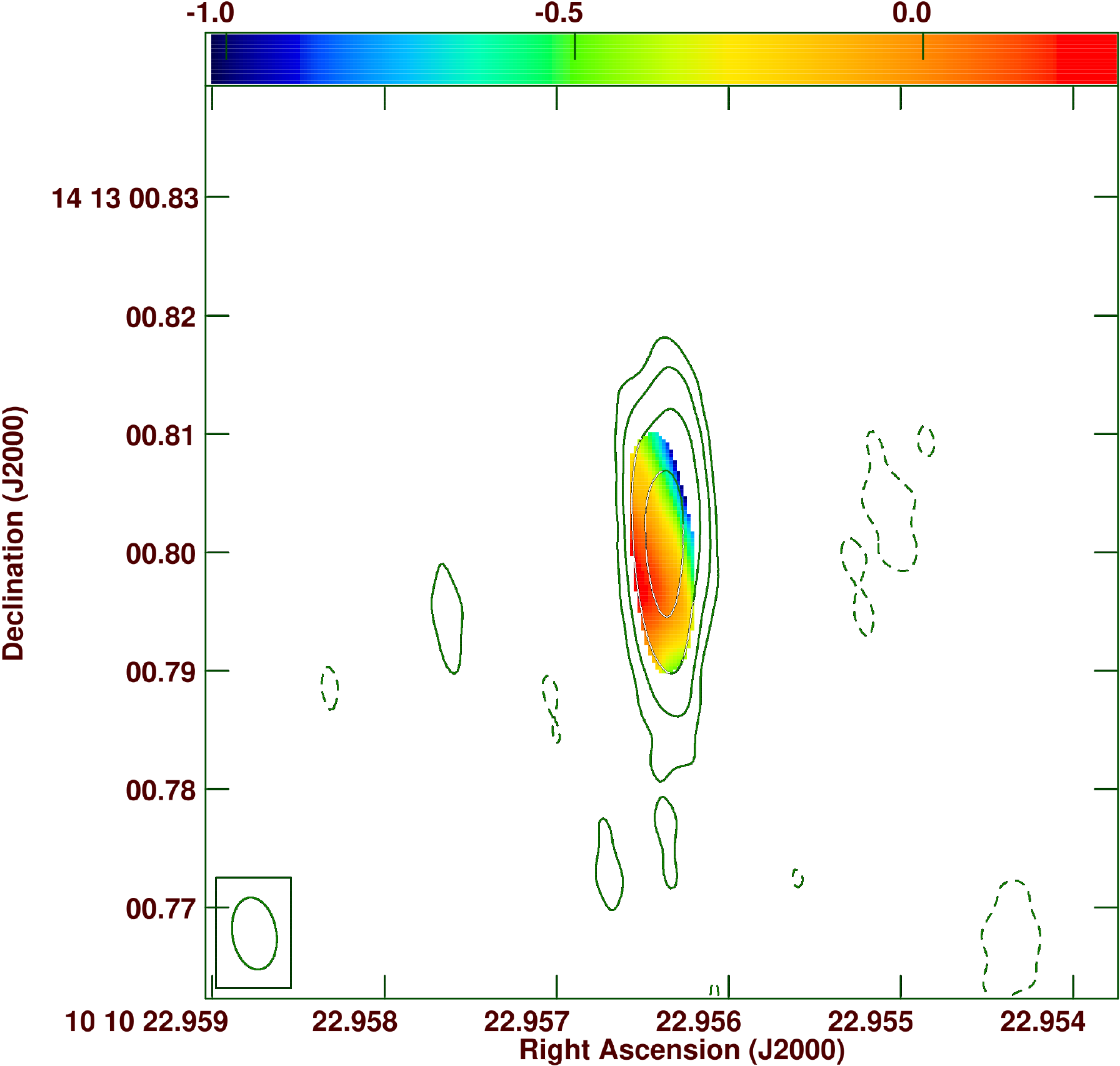}
\caption{Spectral index map of J1010$+$1413 measured between the two observing frequencies, $1.7$ and $4.9$\,GHz (color scale). The contours represent the naturally-weighted $1.7$-GHz image (Figure~\ref{fig:maps}). The restoring beam size (HPBW) is the same as the HPBW of the $1.7$\,GHz map and shown in the lower-left corner.}  \label{fig:spx}
\end{figure}

We present the spectral index map of J1010$+$1413 measured between $1.7$\,GHz and $4.9$\,GHz in Figure~\ref{fig:spx}. We set the same pixel size and map size for the two images, and the larger restoring beam size of the $1.7$\,GHz map was used for both images as the common restoring beam. We combined our individual images taken at the two observing frequencies within two weeks by matching their coordinates in \textsc{AIPS}. We used the \textsc{comb} task by setting the blanking at $3\sigma$ image noise levels for both images. The spectral index ($\alpha$, defined as $S \propto \nu^{\alpha}$) of the source as a whole was also derived from the flux densities of the fitted Gaussian model components (Table~\ref{tab:modelfit}): $\alpha \approx -0.45$.

\section{Multi-band Characteristics of J1010$+$1413}
\label{sec:disc}

\subsection{Mas-scale Radio Structure}
\label{subsec:evn}

With our VLBI observations, we resolved the assumed core component of J1010$+$1413 designated as HR:A in the lower-resolution radio maps of \citet{Jarvis}. The high brightness temperature of this feature ($\sim 10^7$\,K) indicates that the radio emission we detected originates from physical processes associated with AGN activity \citep[e.g.,][]{sfrTb}, in agreement with the finding of \citet{Jarvis2}. Its single-component morphology is consistent with the single AGN core interpretation preferred by \citet{Jarvis}. However, this brightness temperature value still allows that the VLBI component is a hotspot appearing as the consequence of the interaction between the jet and the interstellar medium. Similar brightness temperatures have been measured for terminal hotspots in the jet (e.g., \citealt{hotspotTb2,2010MNRAS.402...87A,hotspotTb}). The spectral index of the feature ($\alpha\approx-0.45$), at around the dividing line between flat and steep spectra, and the extended nature of the emitting region (Figure~\ref{fig:maps}) also make the identification of the component with an AGN core challenging. However, the moderately flat spectrum of the nuclear region can be caused by the blending of the flat-spectrum core and a steep-spectrum jet component, or alternatively explained by multiple episodes of jet activity as unresolved, younger jets/lobes outshine the core or the radio core recently turned off \citep{Jarvis}. Shocked quasar winds in principle may also contribute to the radio emission, causing steeper spectra \citep{steepspectra,Jarvis2}. However, the collimated radio structure seen between the pc and the kpc scales (see below) is incompatible with the quasar wind description.

\subsection{Comparison with Optical Data} 
\label{subsec:shift}

\begin{figure}[h!]
\epsscale{1.1}
\plotone{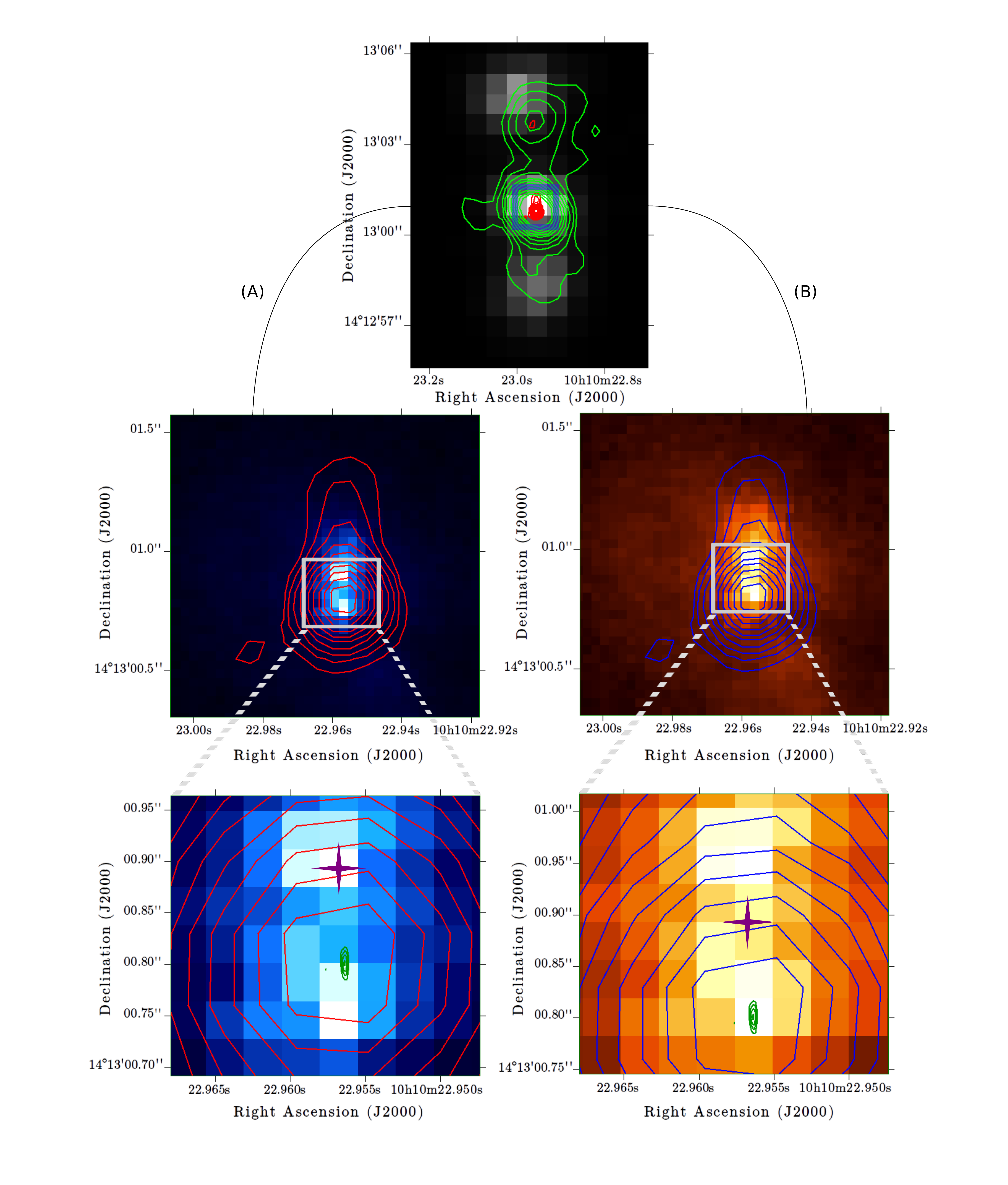}
\caption{Comparison of the archival optical and radio data for J1010$+$1413 with the results of our EVN observations. {\it Upper panel}: Archival VIMOS [\ion{O}{3}] maps with gray scale. The light green and red contours represent the archival low-resolution and high-resolution VLA observations, respectively, taken at $6.2$\,GHz. {\it Middle panels}: Zoom-in images of the upper panel. The color map represents the re-calibrated archival \textit{HST}/WFC3 data with the optical medium band F689M in which the point sources are spatially coincident with the [\ion{O}{3}]-emitting point sources. The maps show the two proposed scenarios (A -- left, B -- right) with different positional shifts as discussed in Section~\ref{subsec:shift}. {\it Bottom panels}: Zoom-in images of the middle panels. The green contours represent our EVN image obtained at $1.7$\,GHz. The lowest contour level is drawn at $4.5\sigma$ image noise. Further contours increase by a factor of 2. The purple star symbol indicates the {\it Gaia} EDR3 position, and its size represents its positional uncertainty. North is up and east is to the left.\label{fig:cmaps}}
\end{figure}

In the upper panel of Figure~\ref{fig:cmaps}, we present archival integral field spectroscopy (IFS) data observed with the VIsible Multi-Object Spectrograph (VIMOS) on  the Very Large Telescope (VLT) operated by the European Southern Observatory (Program ID: 092.B-0062, PI: C. Harrison), together with the results of the low-resolution ($\sim1\arcsec$ beam) and high-resolution ($\sim0\farcs25$ beam) observations of the VLA (Program ID: 13B-127, PI: C. Harrison). The results of these observations were published by \citet{Jarvis} and the data were made publicly available in the Newcastle University's data repository\footnote{\url{https://data.ncl.ac.uk/}}. The optical and radio features are apparently spatially coincident with each other on arcsec scales. Consistently, a close connection is observed between the radio and ionized gas morphologies, according to \citet{Jarvis}. 
To compare the observations taken at different wavelengths on sub-arcsec scales, we downloaded the archival \textit{HST}/WFC3 data (Proposal ID:14730, PI: A. Goulding) from the Barbara A. Mikulski Archive for Space Telescopes\footnote{\url{https://archive.stsci.edu/}} (MAST). The results of these observations were published by \citet{Goulding}. We found that there is a significant positional offset between the \textit{HST}/WFC3 detected optical and the VLBI detected radio emitting regions. The center of the \textit{HST}/WFC3 detected [\ion{O}{3}] emission region is located $\sim345$\,mas to the northeast with respect to the VLBI position of J1010$+$1413. We suspect that this apparent displacement is, at least in part, related to the limited accuracy of the absolute astrometric calibration of the \textit{HST}/WFC3 data \citep{absastro}. 

To tackle the astrometry issue, we turned to the {\it Gaia} Early Data Release 3 (EDR3) \citep{gaia, gaiaedr3}. The {\it Gaia} position of J1010$+$1413 (marked with purple stars in Figure~\ref{fig:cmaps} in the bottom panels) is also offset by a large amount,  $\sim270$\,mas to the southwest from the center of the [\ion{O}{3}] emission region observed by the \textit{HST}/WFC3. The displacement between the EVN position and the {\it Gaia} EDR3 position of J1010$+$1413, usually both known to be accurate at mas level or better, is $95$\,mas (corresponding to $\sim315$\,pc projected linear distance). The {\it Gaia} EDR3 positional uncertainties are $4.0$\,mas in right ascension and $3.3$\,mas in declination. However, the value of the astrometric excess noise parameter is substantial, $\epsilon=25.8$\,mas. Its high significance ($D=431$) indicates that it has to be taken into account when calculating the positional uncertainties. The square root of the sum of squares of the standard error and the astrometric excess noise is $\sim26$\,mas, which is, nevertheless, still below the offset between the \textit{HST}/WFC3 and VLBI positions of J1010$+$1413.

To check and possibly correct the alignment of the different images, we searched for objects in the \textit{HST}/WFC3 image that are also detected in the Sloan Digital Sky Survey (SDSS). Thanks to the wide field of view of WFC3, we found two other nearby objects\footnote{We note that these objects were not detected with \textit{Gaia}.} in the F689M filter image. One of them is located at $\sim7\farcs9$ to the northeast, the other one is at $\sim9\farcs9$ to the southwest from the position of J1010$+$1413. Assuming that both are identifiable with the nearest object found in the SDSS catalog, they are SDSS\,J101023.33+141306.4 and SDSS\,J101022.57+141252.6, respectively, in Data Release 12 \citep[SDSS DR12,][]{sdss-dr12}. Following this identification scheme, we found that both SDSS objects are located southwest of their assumed counterpart in the \textit{HST}/WFC3 image, just like J1010$+$1413 itself. Based on the two nearby field objects, we calculate a shift of $293\pm90$\,mas in the northeastern direction required to make the \textit{HST}/WFC3 source positions consistent with the SDSS astrometry. Applying this shift to the \textit{HST}/WFC3 image, we found that the two distinct \textit{HST}/WFC3 point sources observed by \citet{Goulding} have more consistent positions in comparison with the radio structure and the {\it Gaia} EDR3 position of J1010$+$1413. Although the registration of an \textit{HST}/WFC3 image relative to {\it Gaia} coordinates \citep[e.g.,][]{absastrogaia1} would usually provide more precise calibration, in the absence of those, astrometry relative to SDSS DR12 can also be applied to determine a shift with an acceptable accuracy \citep[e.g.,][]{absastro}. 

The fact that {\it Gaia} observed only one source in the J1010$+$1413 system can be explained by its limited angular resolution. Below separations of $\sim 0\farcs7$, the completeness in close source pairs falls very rapidly \citep{gaiasep}, suggesting that the two \textit{HST}/WFC3 point sources may appear blended into one object in the {\it Gaia} EDR3 catalog. The high (and significant) value of the astrometric excess noise parameter of J1010$+$1413 in the \textit{Gaia} catalogs\footnote{The astrometric excess noise and its significance are lower in the {\it Gaia} DR2 catalog \citep{gaiadr2} than in EDR3, however they are still substantial, $\epsilon_\mathrm{DR2}=11.5$\,mas $D_\mathrm{DR2}=230$.} can be a further indication of this blending effect. The astrometric excess noise describes the disagreement between the observations of a source and the best-fitting standard astrometric model \citep{excessnoise1}\footnote{\url{https://gea.esac.esa.int/archive/documentation/GDR2/Gaia_archive/chap_datamodel/sec_dm_main_tables/ssec_dm_gaia_source.html}}.

In the following, we discuss two plausible scenarios, both being compatible with the positional shift we perform on the F689M image within its uncertainty.

\begin{itemize}

\item[(A)]
We may assume that the {\it Gaia}/EDR3 position of J1010$+$1413 is spatially coincident with the northern one of the \textit{HST}/WFC3 point sources. This case is shown in the middle and lower left panels of Figure~\ref{fig:cmaps} with two different fields of view. By accepting this positional shift of the \textit{HST}/WFC3 image, the feature we detected with the EVN is located close to the southern \textit{HST}/WFC3 point source, $\sim 45\,$mas away from it towards the north. 

\item[(B)]
Perhaps more likely, the {\it Gaia} EDR3 position of J1010$+$1413 may fall between the two \textit{HST}/WFC3 point sources, as they are blended together because of the limited {\it Gaia} resolution. In that case, 
the {\it Gaia} position is located $\sim35$\,mas away from the northern (and, notably, brighter) \textit{HST}/WFC3 point source towards the south, and the southern \textit{HST}/WFC3 point source coincides with the VLBI component. This scenario is depicted on the right-hand side of Figure~\ref{fig:cmaps}.

\end{itemize}

We stress that the positional uncertainties of the SDSS objects which we used to derive the shift of the \textit{HST}/WFC3 image to improve its absolute astrometric registration prevent us from deciding which one of the scenarios mentioned above is more realistic.

\subsection{Kpc-scale Morphology} 
\label{subsec:kpc}

The outflow structure of J1010+1413 was investigated employing integral field spectroscopy by \citet{Harrison}, and long-slit spectroscopy by \citet{Sun}. The multiple kinematic components seen in the emission line profile of J1010$+$1413 can be explained by outflows, rotating gas disks or merging processes. The irregular velocity field and the wide [\ion{O}{3}] emission-line profile (the line width containing 80\% of the flux, $W_{80}\sim1450$\,km\,s$^{-1}$, see \citet{Harrison}, \citet{Sun} or Appendix A in \citealt{Jarvis}) are more likely the signs of outflows or merger remnants than rotating gas disks \citep{Harrison}. \citet{Harrison} also found that the spatially distinct narrow emission-line regions on the [\ion{O}{3}] emission-line peak signal-to-noise ratio maps indicate merger remnants, halo gas or outflow remnants located to the north and south of the nucleus. \citet{Sun} concluded that the wide [\ion{O}{3}] component is an unambiguous sign of a high-velocity outflow and determined its radius of $R_{\mathrm{KDR}}=8$\,kpc. Such a wide emission-line component also indicates an extremely high outflow velocity of $v\sim W_{80}/1.3 = 1110$\,km s$^{-1}$. Thus, the [\ion{O}{3}] clouds are signs of ionization cones or bipolar outflows and these regions are also spatially coincident with the locations of velocity drops found by \citet{Sun}. The presence of sudden velocity drops in the spectroscopic data is a common phenomenon among type 2 AGN and is consistent with the constant-velocity spherical outflow model \citep{outflowmodel}.

\citet{Jarvis} found a close connection between the radio and ionized gas structures, and also concluded that the kpc-scale morphology of J1010$+$1413 is dominated by galactic outflows. The north--south elongated triple radio structure in the low-resolution VLA map indicates core--jet geometry on kpc scales, which is consistent with the presence of jet-driven galactic outflows. These powerful jet-driven outflows provide a plausible interpretation of the kinematically complex structure found by \citet{Harrison}.

The low-resolution radio images of \citet{Jarvis} revealed that both the southern and northern radio lobes are seemingly co-spatial with the [\ion{O}{3}] emission-line regions. \citet{Jarvis} interpreted their findings with a jet both pushing a gas cloud away and also being deflected by it. \citet{3C316} explained the results of their observations of the brightest radio galaxy with double-peaked [\ion{O}{3}] lines, 3C\,316, in a similar way.
The velocity drops in the $W_{80}$ profiles found by \citet{Sun} also indicate interactions between the outflows and the galactic medium, in agreement with the findings of \citet{Jarvis}.

The radio morphology of J1010$+$1413 implies asymmetric structure on kpc scales \citep[see][and the upper panel of Figure~\ref{fig:cmaps}]{Jarvis}. Apparently, the northern lobe is brighter and seen farther away from the central region than the southern one. This can be explained by projection effects \citep{jets}. Adopting this interpretation, the northern lobe, being farther than the southern lobe, implies interaction between the approaching jet and the interstellar medium, while the southern lobe is associated with the receding jet. 
The hotspot labeled as HR:C in the high-resolution radio maps of \citet{Jarvis} may be the reason why the approaching jet looks brighter than the receding one. Additionally, the `median velocity' (i.e. the velocity at $50\%$ of the cumulative flux, $v_{50}$) map of J1010$+$1413 \citep{Jarvis} revealed that the northern regions are strongly blue-shifted. However, the presence of a hotspot suggests environmental asymmetries which may also be responsible for the observed jet structure, thus the radio asymmetry is not necessarily due to projection effects. 

The pc-scale radio morphology revealed by our VLBI imaging observations, especially at 4.9\,GHz (Figure~\ref{fig:maps}) also strengthens the core--jet interpretation if we identify the most compact region as an AGN core component and the northward-elongated structure as the approaching jet. This is also supported by the spectral index map in Figure~\ref{fig:spx} that may indicate a gradually steepening spectrum further away from the core along the jet towards the north.

Similar characteristics such as the triple radio structure roughly aligning with high-velocity outflows and the broad ionized features were found by \citet{mrk231} when investigating the kpc-scale morphology of the nearby quasar, Mrk\,231. The unusual optical--ultraviolet properties of Mrk\,231 suggest an ongoing merger event \citep{mrk231binary}. However, just like in our case, some other characteristics challenge the merging system scenario \citep{mrk231(3),mrk231(2),mrk231(1)}. Interestingly, Mrk\,231 is between the radio-quiet and radio-loud populations, similarly to J1010$+$1413, which is a quasar with modest radio luminosities \citep{Jarvis2}.

Considering these findings, we conclude that the radio morphology of J1010$+$1413 implies a single AGN with jet-driven outflows. The possible coordinate shifts discussed in Section~\ref{subsec:shift} suggest that the jetted AGN is either located between the two \textit{HST}/WFC3 point sources (scenario A) or is coincident with the southern one (scenario B; see also Figure~\ref{fig:cmaps}). However, in Section~\ref{sec:scenarios} we will elaborate on the possible presence of another, radio-undetected AGN in the system, in view of the positional shifts and the kpc-scale morphology.

\section{Single AGN or Dual AGN?} 
\label{sec:scenarios}

\citet{Goulding} suggested the presence of a dual SMBH system in J1010$+$1413 based upon their {\it HST} observations. The two optical point sources could not be resolved by {\it Gaia}. However, the large astrometric excess noise may indicate that the {\it Gaia}-detected single source is a blend of the two optical regions seen by the {\it HST}. In this section, we discuss our new VLBI data together with the information available in the literature in the context of a single obscured AGN or a dual AGN scenario, since the spectroscopic observations allow any of these interpretations.

\subsection{Single Obscured AGN}
\label{subsec:single-agn}

The type 2 AGN classification of J1010$+$1413 \citep{type2}, supported by the lack of broad lines in the optical spectrum and its location in the \textit{WISE} color--color diagram \citep{wisecolorcolor}, suggests heavy dust obscuration in the system. In general, type 2 AGN are seen edge-on and obscured by a geometrically thick toroidal structure (i.e., the dust torus). However, larger-scale structures (e.g., dust lanes) in the host galaxy can also result in obscuration \citep[e.g.,][]{viewangle}. In the case of J1010$+$1413, the other observed properties can be reconciled with this picture as follows.

\begin{itemize}

\item
The two closely-separated ($\sim 430$\,pc) optical continuum sources coincident with the [\ion{O}{3}] emitting point sources \citep{Goulding} correspond to two opposite (the northern and southern) sides of the NLR in a single AGN. The central region is obscured from our view in the optical by a dusty structure that is extended to $\sim 400$\,pc, i.e. a dust lane rather than a smaller-scale torus which is believed to be $\approx 0.1 - 10$\,pc thick \citep{torussize}. In other words, the emission of the single NLR splits into two parts, and the two \textit{HST}/WFC3 point sources observed by \citet{Goulding} are related to the same single AGN. 

\item
Multi-wavelength investigations of the torus revealed a multiphase, multicomponent dusty molecular environment rather than a single entity \citep{torussize, torus}. In this picture, additional dust components play a role, providing another explanation of the heavy obscuration. Dusty molecular winds can result in some additional obscuring material to the torus as the molecular gas flows in from galactic scales of $\sim 100$\,pc to the sub-pc environment via a disk \citep{torus,torus2}. This interpretation explains the larger-scale obscuration and also the relatively steep radio spectrum of the nuclear region since shocked quasar winds are one of the plausible explanations of the unusual spectral properties of J1010$+$1413. However, it is in tension with the collimated nature of the radio jets observed at pc and kpc scales. 

\item
The radio emission is unaffected by dust obscuration. It is therefore reasonable to assume that the compact radio feature we detect with the EVN and e-MERLIN (Figure~\ref{fig:maps}) is either the optically obscured, jetted AGN itself, located between the divided parts of the NLR (scenario A described in Section~\ref{subsec:shift} and shown on the left-hand side in Figure~\ref{fig:cmaps}), or perhaps a hotspot in the jet launched by the central AGN whose weak radio core remains undetected in our high-resolution imaging observations (scenario B shown on the right-hand side in Figure~\ref{fig:cmaps}). Unfortunately, the lack of sufficiently accurate astrometric registration between the optical and radio imaging prevents us from determining the exact location of the VLBI-detected radio feature with respect to the optical point sources. 

\item
The two-sided arcsec-scale radio structure is a natural consequence of jet activity in the central AGN. The dominance of the northern side in both the low- and high-resolution VLA images (Figure~\ref{fig:cmaps}), together with the core--jet interpretation of our dual-frequency VLBI images (Figures~\ref{fig:maps} and \ref{fig:spx}) is straightforwardly explained with a classical double-lobed radio structure oriented close to the plane of the sky. The inclination of such a structure makes the radio emission on the approaching (northern) side appear brighter.  

\item
The idea that the dust lane splits the NLR into two parts implies an edge-on host galaxy. The continuity between the dust in the pc-scale torus and the kpc-scale dust lane found by \citet{viewangle} is consistent with this picture, since the type 2 AGN classification of J1010$+$1413 requires the same viewing angle (i.e., perpendicular to the radio jet axis).

\end{itemize}

We note that the morphological analysis made by \citet{Goulding} does not support the presence of an obscuring dust lane in the system.

\subsection{Dual AGN}
\label{subsec:dual-agn}

How do the observations of J1010$+$1413 at multiple wavebands fit the dual AGN scenario proposed by \citet{Goulding}?

\begin{itemize}

\item
The two optical stellar-continuum-only and [\ion{O}{3}] emitting point sources with $0\farcs13$ angular separation ($\sim 430$\,pc projected linear separation) may suggest the presence of two distinct AGN. The luminous [\ion{O}{3}] point sources are spatially coincident with stellar components, indicating two independent NLRs in the system (e.g., \citealt{stellarcomp}).

\item
Double-peaked optical narrow emission lines may also indicate the presence of two accreting AGN with different radial velocity components (e.g., \citealt{sdssdualagn}).

\item
The VLBI observations show that one of the optically detected components in the dual AGN system shows jet activity. According to our scenario B (Section~\ref{subsec:shift}; right-hand side in Figure~\ref{fig:cmaps}), this must be the southern AGN. Its position in the \textit{HST}/WFC3 image can be shifted so that it matches with the VLBI radio position within the astrometric uncertainties. In this case, the accurate \textit{Gaia} position falls between the two sources resolved in the \textit{HST}/WFC3 image, suggesting that \textit{Gaia} in fact determines a position for the blended optical AGN sources. 

\item
The radio structure seen on multiple scales and extended up to several arcsecs  \citep{Jarvis} is most likely associated with the jetted component of the dual AGN system. The explanation of its orientation given in the single AGN scenario above is also valid in this case for the radio-emitting AGN of the assumed dual AGN system.

\item
The type 2 AGN classification of J1010$+$1413 \citep{type2} is more difficult to reconcile with the dual AGN since at least one component in the system should be obscured by a large column of dust. It is therefore unexpected to be detected as an optical point source in the  \textit{HST}/WFC3 image. However, we note that \citet{Goulding} identified the two \textit{HST}/WFC3 point sources as [\ion{O}{3}]-luminous NLRs. Because NLRs are extended, their emission may not be completely obscured, which makes it possible to observe them using optical spectroscopic techniques (see \citealt{obsagn} and references therein). 

\end{itemize}

\section{Conclusion} \label{sec:concl}
We presented the results of our EVN and e-MERLIN observations of the promising sub-kpc separation dual AGN candidate J1010$+$1413. With our high-resolution VLBI observations, we resolved the assumed core region of the object. Although earlier \textit{HST} observations suggest the presence of two AGN in that region, only a single feature can be seen in our VLBI maps taken at $1.7$\,GHz and $4.9$\,GHz. Regardless of its AGN core or hotspot identification, our observations imply the presence of at least one AGN in this system. Due to the lack of the precise absolute astrometric calibration of the \textit{HST}/WFC3 data, it is uncertain whether the feature we detected can be identified with one of the point sources found by the \textit{HST}. Nevertheless, we derived a shift of $293\pm90$\,mas of the \textit{HST} images to the northeastern direction, which leads to two plausible interpretations of the nature of J1010+1413. Both scenarios are consistent with the shift within the errors and both can explain the presence of double-peaked optical emission lines in the spectra, hence we collected some other observed properties of the object to strengthen or falsify the proposed dual AGN identification of J1010$+$1413. In the single obscured AGN scenario, the feature we detected is a jetted AGN or a hotspot in the jet located between the two \textit{HST}/WFC3 point sources. The idea that the two \textit{HST}/WFC3 point sources are the two parts of the same NLR divided by an obscuring dust lane explains the optical spectroscopic properties of the object and also consistent with its type 2 AGN classification. Our mas-scale EVN observations are consistent with this picture, and the two-sided arcsec-scale radio structure revealed by \cite{Jarvis} also suggests the presence of a single AGN in the system with outflows. However, the morphological analysis made by \cite{Goulding} indicates two AGN with sub-kpc separation. Here the southern \textit{HST}/WFC3 point source would be spatially coincident with the feature we detected, consequently the northern optical point source is a radio-quiet AGN which is not visible in our high-resolution VLBI maps.
As shown above, both scenarios can explain most of the observed properties of J1010+1413, consequently the earlier dual AGN classification requires further observational evidence. Hard X-ray observations might shed light on the nature of the system by constraining the absorbing column density and revealing the X-ray emission from the single or dual AGN. However, due to the limited resolution of hard X-ray observations, it seems to be challenging even in the future.

\acknowledgements
The EVN is a joint facility of independent European, African, Asian and North American radio astronomy institutes. The e-MERLIN is a National Facility operated by the University of Manchester at Jodrell Bank Observatory on behalf of STFC. Scientific results from data presented in this publication are derived from the following EVN project code: EG109. The research leading to these results has received funding from the European Commission Horizon 2020 Research and Innovation Programme under grant agreement No. 730562 (RadioNet). We thank the Hungarian National Research, Development and Innovation Office (OTKA K134213, 2018-2.1.14-T\'ET-CN-2018-00001) for support. E.K. was supported by the Premium Post-doctoral Research Program of the Hungarian Academy of Sciences. This work has made use of data from the European Space Agency (ESA) mission Gaia (https://www.cosmos.esa.int/gaia), processed by the Gaia Data Processing and Analysis Consortium (DPAC, https://www.cosmos.esa.int/web/gaia/dpac/consortium). Funding for the DPAC has been provided by national institutions, in particular the institutions participating in the Gaia Multilateral Agreement.



\begin{thebibliography}{22}

\bibitem[Alam et al.(2015)]{sdss-dr12} Alam, S., Albareti, F.~D., Allende Prieto, C., et al.\ 2015, \apjs, 219, 12. doi:10.1088/0067-0049/219/1/12

\bibitem[Alef \& Porcas(1986)]{selfcal} Alef, W., \& Porcas, R. W.\ 1986, \aap, 168, 365

\bibitem[Alonso-Herrero et al.(2021)]{torus2} Alonso-Herrero, A., Garc{\'\i}a-Burillo, S., Hoenig, S.~F., et al.\ 2021, arXiv:2107.00244

\bibitem[An et al.(2010)]{2010MNRAS.402...87A} An, T., Hong, X.~Y., Hardcastle, M.~J., et al.\ 2010, \mnras, 402, 87. doi:10.1111/j.1365-2966.2009.15899.x

\bibitem[An et al.(2013)]{3C316} An, T., Paragi, Z., Frey, S., et al.\ 2013, \mnras, 433, 1161. doi:10.1093/mnras/stt801

\bibitem[An et al.(2018)]{dual-radio} An, T., Mohan, P., \& Frey, S.\ 2018, Radio Sci., 53, 1211.
doi:10.1029/2018RS006647

\bibitem[Begelman et al.(1980)]{GW} Begelman, M.~C., Blandford, R.~D., \& Rees, M.~J.\ 1980, \nat, 287, 307. doi:10.1038/287307a0

\bibitem[Charlot et al.(2020)]{icrf} Charlot, P., Jacobs, C.~S., Gordon, D., et al.\ 2020, \aap, 644, A159. doi:10.1051/0004-6361/202038368

\bibitem[Chatterjee et al.(2004)]{Chatterjee} Chatterjee, S., Cordes, J.~M., Vlemmings, W.~H.~T., et al.\ 2004, \apj, 604, 339. doi:10.1086/381748

\bibitem[Condon et al.(1982)]{Tb} Condon, J.~J., Condon, M.~A., Gisler, G., et al.\ 1982, \apj, 252, 102. doi:10.1086/159538

\bibitem[Condon(1992)]{sfrTb} Condon, J.~J.\ 1992, \araa, 30, 575. doi:10.1146/annurev.aa.30.090192.003043

\bibitem[De Rosa et al.(2019)]{multi-messenger} De Rosa, A., Vignali, C., Bogdanovi{\'c}, T., et al.\ 2019, \nar, 86, 101525. doi:10.1016/j.newar.2020.101525

\bibitem[Deane et al.(2014)]{deane-triple} Deane, R. P., Paragi, Z., Jarvis, M. J., el al.\ 2014, \nat, 511, 57.
doi:10.1038/nature13454

\bibitem[Fabricius et al.(2021)]{gaiasep} Fabricius, C., Luri, X., Arenou, F., et al.\ 2021, \aap, 649, A5. doi:10.1051/0004-6361/202039834

\bibitem[Fomalont(1999)]{Fomalont} Fomalont, E.~B.\ 1999, in Synthesis Imaging in Radio Astronomy II, ed. G.~B. Taylor, C.~L. Carilli, \& R.~A. Perley, ASP Conf. Ser. 180 (San Francisco: ASP), 301

\bibitem[Fu et al.(2011a)]{double-peaked} Fu, H., Myers, A.~D., Djorgovski, S.~G., et al.\ 2011a, \apj, 733, 103.
doi:10.1088/0004-637X/733/2/103

\bibitem[Fu et al.(2011b)]{sdssdualagn} Fu, H., Zhang, Z.-Y., Assef, R.~J., et al.\ 2011b, \apjl, 740, L44. doi:10.1088/2041-8205/740/2/L44

\bibitem[Gaia Collaboration et al.(2016)]{gaia} Gaia Collaboration, Prusti, T., de Bruijne, J.~H.~J., et al.\ 2016, \aap, 595, A1. doi:10.1051/0004-6361/201629272

\bibitem[Gaia Collaboration et al.(2018)]{gaiadr2} Gaia Collaboration, Brown, A.~G.~A., Vallenari, A., et al.\ 2018, \aap, 616, A1. doi:10.1051/0004-6361/201833051

\bibitem[Gaia Collaboration et al.(2021)]{gaiaedr3} Gaia Collaboration, Brown, A.~G.~A., Vallenari, A., et al.\ 2021, \aap, 649, A1. doi:10.1051/0004-6361/202039657 

\bibitem[Goulding et al.(2019)]{Goulding} Goulding, A.~D., Pardo, K., Greene, J.~E., et al.\ 2019, \apjl, 879, L21. doi:10.3847/2041-8213/ab2a14

\bibitem[Greisen(2003)]{aips} Greisen, E.~W.\ 2003, in Information Handling in Astronomy -- Historical Vistas, ed. A. Heck, Astrophys. Space Sci. Library, 285 (Dordrecht: Kluwer), 109. doi:10.1007/0-306-48080-8$\_$7

\bibitem[Gurvits et al.(1997)]{hotspotTb2} Gurvits, L.~I., Schilizzi, R.~T., Miley, G.~K., et al.\ 1997, \aap, 318, 11

\bibitem[Harrison et al.(2014)]{Harrison} Harrison, C.~M., Alexander, D.~M., Mullaney, J.~R., et al.\ 2014, \mnras, 441, 3306. doi:10.1093/mnras/stu515

\bibitem[Hickox \& Alexander(2018)]{obsagn} Hickox, R.~C. \& Alexander, D.~M.\ 2018, \araa, 56, 625. doi:10.1146/annurev-astro-081817-051803

\bibitem[H{\"o}gbom(1974)]{hybridmapping} H{\"o}gbom, J.~A.\ 1974, \aaps, 15, 417

\bibitem[H{\"o}nig(2019)]{torus} H{\"o}nig, S.~F.\ 2019, \apj, 884, 171. doi:10.3847/1538-4357/ab4591

\bibitem[Hudson et al.(2006)]{3C75(2)} Hudson, D.~S., Reiprich, T.~H., Clarke, T.~E., et al.\ 2006, \aap, 453, 433. doi:10.1051/0004-6361:20064955

\bibitem[Ivezi{\'c} et al.(2002)]{radio-loud} Ivezi{\'c}, {\v{Z}}., Menou, K., Knapp, G.~R., et al.\ 2002, \aj, 124, 2364. doi:10.1086/34406

\bibitem[Jaiswal et al.(2019)]{hotspotTb} Jaiswal, S., Mohan, P., An, T., et al.\ 2019, \apj, 873, 11. doi:10.3847/1538-4357/ab0176

\bibitem[Jarvis et al.(2019)]{Jarvis} Jarvis, M.~E., Harrison, C.~M., Thomson, A.~P., et al.\ 2019, \mnras, 485, 2710. doi:10.1093/mnras/stz556

\bibitem[Jarvis et al.(2021)]{Jarvis2} Jarvis, M.~E., Harrison, C.~M., Mainieri, V., et al.\ 2021, \mnras, 503, 1780. doi:10.1093/mnras/stab549

\bibitem[Karamanavis et al.(2016)]{fluxerror} Karamanavis, V., Fuhrmann, L., Krichbaum, T.~P., et al.\ 2016, \aap, 586, A60. doi:10.1051/0004-6361/201527225

\bibitem[Keimpema et al.(2015)]{JIVEcorrelator} Keimpema, A., Kettenis, M.~M., Pogrebenko, S.~V., et al.\ 2015, Exp. Astron., 39, 259. doi:10.1007/s10686-015-9446-1

\bibitem[King et al.(2011)]{outflowmodel} King, A.~R., Zubovas, K., \& Power, C.\ 2011, \mnras, 415, L6. doi:10.1111/j.1745-3933.2011.01067.x

\bibitem[Koekemoer et al.(2005)]{absastro} Koekemoer, A.~M., McLean, B., McMaster, M., \& Jenkner, H.\ 2005, Demonstration of a Significant Improvement in the Astrometric Accuracy of HST Data, Instrument Science Report ACS 2005-06 (Baltimore: STScI)

\bibitem[Kormendy \& Richstone(1995)]{binaries} Kormendy, J. \& Richstone, D.\ 1995, \araa, 33, 581. doi:10.1146/annurev.aa.33.090195.003053

\bibitem[Leighly et al.(2016)]{mrk231(3)} Leighly, K.~M., Terndrup, D.~M., Gallagher, S.~C., et al.\ 2016, \apj, 829, 4. doi:10.3847/0004-637X/829/1/4

\bibitem[Lindegren et al.(2012)]{excessnoise1} Lindegren, L., Lammers, U., Hobbs, D., et al.\ 2012, \aap, 538, A78. doi:10.1051/0004-6361/201117905

\bibitem[Longair \& Riley(1979)]{jets} Longair, M.~S. \& Riley, J.~M.\ 1979, \mnras, 188, 625. doi:10.1093/mnras/188.3.625

\bibitem[Mullaney et al.(2013)]{type2} Mullaney, J.~R., Alexander, D.~M., Fine, S., et al.\ 2013, \mnras, 433, 622. doi:10.1093/mnras/stt751

\bibitem[M{\"u}ller-S{\'a}nchez et al.(2016)]{nlr} M{\"u}ller-S{\'a}nchez, F., Comerford, J., Stern, D., et al.\ 2016, \apj, 830, 50. doi:10.3847/0004-637X/830/1/50

\bibitem[Nims et al.(2015)]{steepspectra} Nims, J., Quataert, E., \& Faucher-Gigu{\`e}re, C.-A.\ 2015, \mnras, 447, 3612. doi:10.1093/mnras/stu2648

\bibitem[Owen et al.(1985)]{3C75} Owen, F.~N., O'Dea, C.~P., Inoue, M., et al.\ 1985, \apjl, 294, L85. doi:10.1086/184514

\bibitem[Pearson(1995)]{modelfit} Pearson, T. J.\ 1995, in Very Long Baseline Interferometry and the VLBA, ed. J. A. Zensus, P. J.Diamond, \& P. J. Napier, ASP Conf. Ser. 82 (San Francisco: ASP), 267

\bibitem[Ram{\'\i}rez et al.(2017)]{viewangle} Ram{\'\i}rez, E.~A., Aretxaga, I., Tadhunter, C.~N., et al.\ 2017, Front. Astron. Space Sci., 4, 52. doi:10.3389/fspas.2017.00052

\bibitem[Ramos Almeida \& Ricci(2017)]{torussize} Ramos Almeida, C. \& Ricci, C.\ 2017, Nat. Astron., 1, 679. doi:10.1038/s41550-017-0232-z

\bibitem[Richard et al.(2021)]{absastrogaia1} Richard, J., Claeyssens, A., Lagattuta, D.~J., et al.\ 2021, \aap, 646, A83. 
doi: 10.1051/0004-6361/202039462

\bibitem[\protect\citeauthoryear{Rioja et al.}{2017}]{rioja} Rioja M.~J., Dodson R., Orosz G., Imai H., Frey S., 2017, AJ, 153, 105. doi:10.3847/1538-3881/153/3/105

\bibitem[Schwab \& Cotton(1983)]{fringefit} Schwab, F. R., \& Cotton, W. D.\ 1983, \aj, 88, 688
doi: 10.1086/113360
 
\bibitem[Sesana et al.(2008)]{PTA} Sesana, A., Vecchio, A., \& Colacino, C.~N.\ 2008, \mnras, 390, 192. doi:10.1111/j.1365-2966.2008.13682.x

\bibitem[Shen et al.(2011)]{stellarcomp} Shen, Y., Liu, X., Greene, J.~E., et al.\ 2011, \apj, 735, 48. doi:10.1088/0004-637X/735/1/48

\bibitem[Shepherd et al.(1994)]{difmap} Shepherd, M.~C., Pearson, T.~J., \& Taylor, G.~B.\ 1994, \baas, 26, 987

\bibitem[Shupe et al.(1998)]{redshift} Shupe, D.~L., Fang, F., Hacking, P.~B., et al.\ 1998, \apj, 501, 597. doi:10.1086/305825

\bibitem[Silpa et al.(2021)]{mrk231(1)} Silpa, S., Kharb, P., O'Dea, C.~P., et al.\ 2021, \mnras, 507, 2550. doi:10.1093/mnras/stab2110

\bibitem[Sokolovsky et al.(2011)]{pref-jet} Sokolovsky, K. V., Kovalev, Y. Y., Pushkarev, A. B., \& Lobanov, A. P.\ 2011, \aap, 532, A38.
doi: 10.1051/0004-6361/201016072

\bibitem[Sun et al.(2017)]{Sun} Sun, A.-L., Greene, J.~E., \& Zakamska, N.~L.\ 2017, \apj, 835, 222. doi:10.3847/1538-4357/835/2/222

\bibitem[Veilleux et al.(2016)]{mrk231(2)} Veilleux, S., Mel{\'e}ndez, M., Tripp, T.~M., et al.\ 2016, \apj, 825, 42. doi:10.3847/0004-637X/825/1/42

\bibitem[Wang et al.(2021)]{mrk231} Wang, A., An, T., Jaiswal, S., et al.\ 2021, \mnras, 504, 3823. doi:10.1093/mnras/stab587

\bibitem[Wright(2006)]{cosmology} Wright, E.~L.\ 2006, \pasp, 118, 1711. doi:10.1086/510102

\bibitem[Wright et al.(2010)]{wisecolorcolor} Wright, E.~L., Eisenhardt, P.~R.~M., Mainzer, A.~K., et al.\ 2010, \aj, 140, 1868. doi:10.1088/0004-6256/140/6/1868

\bibitem[Yan et al.(2015)]{mrk231binary} Yan, C.-S., Lu, Y., Dai, X., et al.\ 2015, \apj, 809, 117. doi:10.1088/0004-637X/809/2/117



\end{thebibliography}
\end{document}